\documentclass[
reprint,
superscriptaddress,
showpacs,preprintnumbers,
nobibnotes,
 amsmath,amssymb,
aps,
]{revtex4-1}

\usepackage{graphicx}
\usepackage{dcolumn}
\usepackage{bm}
\usepackage[colorlinks,linkcolor=blue,citecolor=blue,urlcolor=blue,hyperindex,driverfallback=dvipdfm]{hyperref}
\usepackage[mathlines]{lineno}

\begin{document}

\title{Tailoring reflection of graphene plasmons by focused ion beams}

\affiliation{The Key Laboratory of Weak-Light Nonlinear Photonics, Ministry of Education, School of Physics and TEDA Applied Physics Institute, Nankai University, Tianjin 300457, China}

\author{Weiwei Luo}
\author{Wei Cai}
\email{weicai@nankai.edu.cn}
\author{Wei Wu}
\author{Yinxiao Xiang}
\author{Mengxin Ren}
\author{Xinzheng Zhang}
\author{Jingjun Xu}
\email{jjxu@nankai.edu.cn}

\date{\today}

\begin{abstract}
Graphene plasmons are of remarkable features that make graphene plasmon elements promising for applications to integrated photonic devices. The fabrication of graphene plasmon components and control over plasmon propagating are of fundamental important. Through near-field plasmon imaging, we demonstrate controllable modifying of the reflection of graphene plasmon at boundaries etched by ion beams. Moreover,  by varying ion dose at a proper value, nature like reflection boundary can be obtained. We also investigate the influence of ion beam incident angle on plasmon reflection. To illustrate the application of ion beam etching, a simple graphene wedge-shape plasmon structure is fabricated and performs excellently, proving this technology as a simple and efficient tool for controlling graphene plasmons.

\end{abstract}


\maketitle

\section{\label{sec:level1}INTRODUCTION}

Graphene plasmons, possessing unique and fascinating properties \cite{NGM2004,KCG2011,JBS2009,JGH2011}, like electrostatic tunability, extremely high field confinement and less intrinsic loss, are promising in integrated photonic applications \cite{GPN2012,FTS2013,VE2011,RLJ2015}, such as sensors, waveguides or modulators. Various aspects of graphene plasmons, ranging from methods of launching and detecting \cite{CBA2012,FRA2012,APA2014} to reducing the propagation loss \cite{WLG2014} have been studied experimentally with the aid of near-field optical microscope. Moreover, the control over the propagation of graphene plasmons \cite{VE2011,APA2014} is of particular importance in integrated device applications, wherein the ability to manage reflection of plasmon waves is fundamental. Graphene plasmons have been proved to reflect efficiently at graphene edges \cite{CBA2012,FRA2012}, grain boundaries \cite{FRG2013} and nano gaps between SiC terraces \cite{CNN2013}. Solving how to realize guided plasmon reflection at any required nanozone of graphene is essential, while all the present acquired reflection boundaries emerge naturally. Therefore, the introduction of controllable artificial defects on graphene for plasmon reflection is necessary.

Electron beam lithography (EBL) followed by plasma etching is widely used for patterning graphene nanostructures \cite{YLL2012,YLC2012,YLZ2013,BJS2014,RLJ2015} with spatial resolution of tens of nanometers. Nevertheless, the fabrication process is onerous and commonly results in resists remain \cite{SLL2013}. Besides, the etching of graphene relying on plasma results in totally removed of carbon atoms. Plasmon structures of different reflection characteristics are unachievable.

In this letter, we prove the controllable plasmon reflection boundaries fabricated by focused ion beams (FIB). FIB has been proposed as an efficient tool for graphene modification \cite{KB2007,BLS2009,LBW2009,ZHS2014,KBL2015}, with the advantages of process-simple and high spatial resolution. However, the plasmon properties of FIB etched graphene are still unrevealed.  In this work, through scattering-type scanning near-field optical microscope (s-SNOM), plasmon reflection is observed near graphene boundaries formed by FIB. By controlling the exposure dose, graphene defect structures with varied plasmon reflection properties are obtained, and fabricated reflection boundary comparable to natural one can be realized. A simple wedge-shape graphene structure is manufactured and exhibits well-behaved plasmon properties, demonstrating FIB as a flexible and powerful tool for the realization of graphene plasmonic devices. At last, we exhibit that optimized results can be achieved by varying ion beam incident direction.

\section{\label{sec:level2}Experiment and Results}
\begin{figure}[htb]
\centerline{\includegraphics[width=8.5cm]{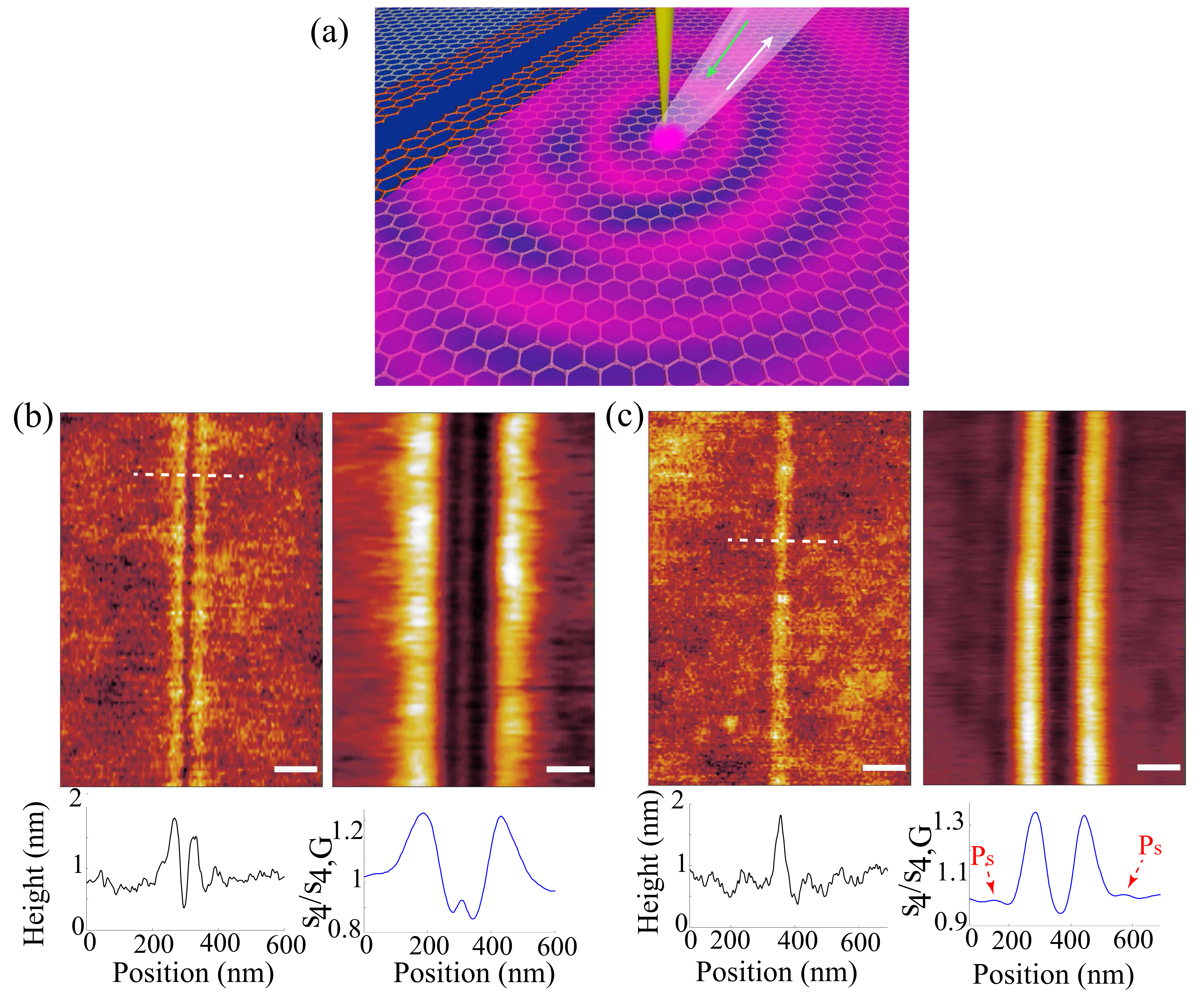}}
\caption{ Nano-imaging of FIB-fabricated graphene structures with s-SNOM. (a) Illustration of plasmon imaging principle. Mid-infrared laser is focused on the metallic AFM tip, which then launches surface plasmon waves in graphene. Propagating plasmon waves are partially reflected back by the defects caused by FIB fabrication,  interfering with the launched waves. (b)-(c) AFM topography (left) and near-field optical images (right, at $\lambda_0$ of 10.653 $\mu$m) of FIB etched graphene. Ion dwell times are (b) 100 $\mu$s  and (c) 5 $\mu$s. Bottom curves in (b) and (c) show line profiles taken along the corresponding dashed lines. Scale bars, 100~nm.
}
\label{fig1}
\end{figure}

A single graphene layer was obtained by mechanical cleavage from bulk graphite and transferred to Si substrate with 300 nm thick SiO$_2$ layer. The layer of graphene was discerned under optical microscope and confirmed by micro-Raman spectrum. FIB etching was completed with FEI Helios Nanolab 600i system, employing 30 kV Ga$^+$ with an ion current of 2.56 pA. To investigate the influence of exposure dose on etching results, varied ion dwell times were adopted, ranging from 100 $\mu$s to 1 $\mu$s. Infrared near-field measurements were performed with s-SNOM (Neaspec). The principle of s-SNOM is illustrated in Fig. \ref{fig1}(a). To probe the properties of graphene plasmons, the free-space wavelength of incident infrared laser $\lambda_0$ was tuned from 10.195 to 10.653 $\mu$m. The metallic AFM tip oscillated at 266 kHz with an amplitude of about 60 nm, then near-filed images were acquired by demodulating the backscattering signals at higher harmonics. The fourth harmonics scattering signal s$_4$ is adopted in the study.

Fig. \ref{fig1}(b) and (c) exhibit the simultaneously obtained AFM topographies and near-field images near graphene defect lines, which were etched at different ion beam dwell times, 100 and 5~$\mu$s, respectively. As previously reported \cite{KBL2015}, there exists a threshold ion dose for cutting graphene, below which only amorphization occur. The AFM height profiles at bottom of Fig. \ref{fig1} show that for high dwell time of 100~$\mu$s, ion bombardment causes a deep etching on graphene, with width of about 30 nm, corresponding to carbon atoms sputtered away. In close proximity to graphene cutted zone, amorphization of graphene occurs as ion amount is less. At lower exposure dose with dwell time of 5~$\mu$s, no cutting of graphene is observed because of insufficient exposure of Ga$^+$ ions.

Near the line hole etched at high dose and at the location of lower dose irradiated zone, hillocks of about 1 nm height and 50 nm width emerge. The hillocks stem from the binding of hydrogen atoms and other chemical groups with defected carbon atoms because  of increased reactivity of defects \cite{DSL2004,BK2008,CLN2009,BKK2010,LBS2013}. Near field optical images in Fig. \ref{fig1}(b) and (c) reveal the interference patterns of plasmon waves at both sides of the etched boundaries. For high irradiation time, only one broad peak can be distinguished, indicating large plasmon damping near etched zone, while in Fig. \ref{fig1}(c), the second weaker peaks $P_s$ are observed.

\begin{figure}[htb]
\centerline{\includegraphics[width=8.5cm]{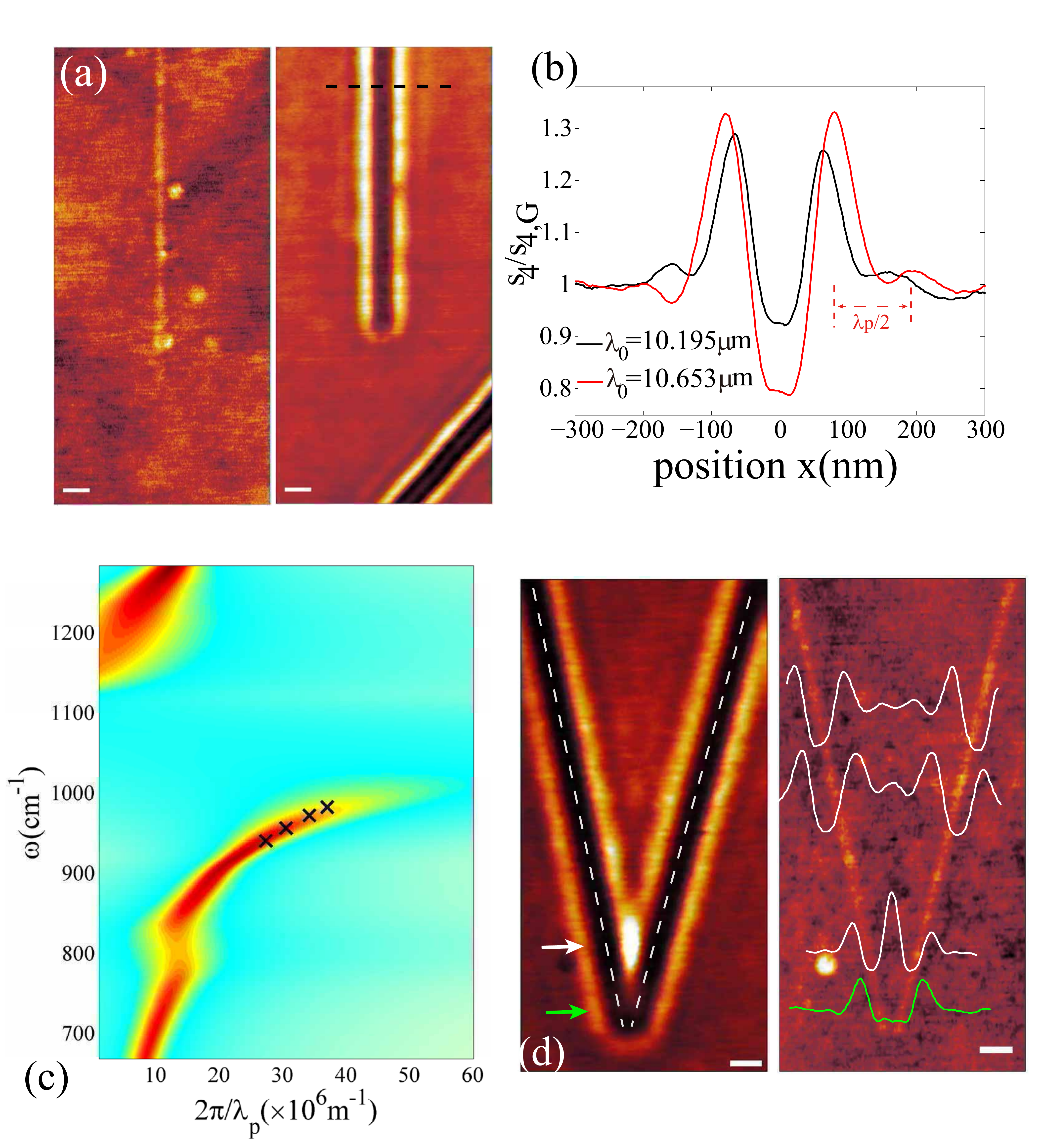}}
\caption{ The plasmon dispersion and near-field imaging of etched graphene at dwell time of 5~$\mu$s. (a) AFM (left) and near-field imaging (right) of FIB etched line defect and natural defect. (b) Line profiles of $s_4$ along the dashed line in (a) at different incident wavelengths. (c) Dispersion of graphene plasmon waves at the Fermi energy of 0.5~eV. Black crosses show the experimental extracted data. (d) Near-field imaging of graphene with two tilted line defects etched by FIB at 10.653~$\mu$m. The two dashed lines indicate the location of etched lines. Scale bars, 100 nm.}
\label{fig2}
\end{figure}

To further verify the observed graphene plasmon reflection, different infrared laser frequencies are applied near the boundary that etched at dwell time of 5 $\mu$s, as shown in Fig. \ref{fig2}(a). Line profiles at 10.195 and 10.653 $\mu$m in Fig. \ref{fig2}(b) illustrate the position out-shifts of first plasmon peaks, implying plasmon dispersion.  Plasmon wavelength $\lambda_p$ can be extracted as twice the distance of  first two interference peaks. The acquired plasmon wave momentums at different frequencies are plotted in Fig. \ref{fig2}(c). Experimental acquired data match well with the dispersion of graphene surface plasmon for Fermi energy of 0.5~eV, verifying reflection of plasmon near ion etched boundaries. Besides, at the bottom of near-field image in Fig. \ref{fig2}(a), both sides of fringes reveal the existence of a natural graphene boundary which cannot be identified from the topography image \cite{FRG2013}. We notice that the Fermi energy of 0.5~eV is larger than 0.4~eV seen in before studies of graphene on silicon dioxide\cite{FRA2012,FRG2013}, which we attribute to the doping effect of Ga$^+$ ions\cite{ZLW2010}.

To illustrate the properties of plasmon reflection at ion etched graphene boundaries, two tilted line defects were drawn by ion beams, indicated by dashed lines in the near-field image in Fig. \ref{fig2} (d). Similar phenomenon is observed as that of tapered graphene ribbon \cite{CBA2012}. In the wider part, plasmon waves reflected from both sides interference with each other, and the overlapping of two second plasmon peaks can be observed. Then in the narrow part, localized plasmon resonances arise, exhibiting enhanced near-field signals at the center of the gap, indicating enhanced local density of optical states. The lowest mode where field maxima at boundaries can also be distinguished, indicated by  green arrow in Fig. \ref{fig2}(d), demonstrating excellent plasmon properties of ion beam etched graphene structure.

\begin{figure}[htb]
\centerline{\includegraphics[width=7cm]{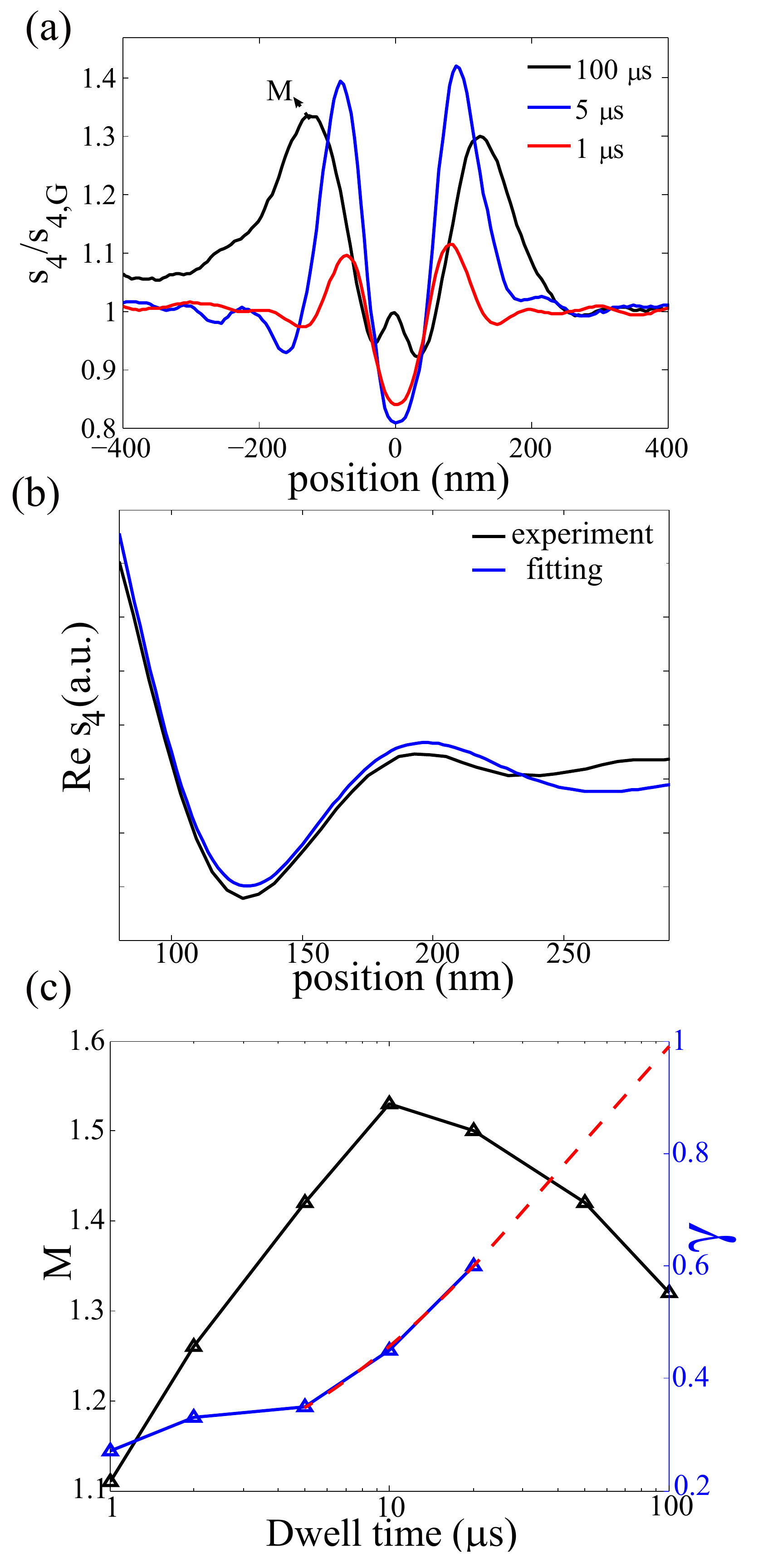}}
\caption{Properties of plasmon propagation and  reflection with varied exposure doses, launched  at $\lambda_0$=10.653~$\mu$m. (a) Near-field signals at different dwell times, all normalized to those far from defects. Value of the first plasmon interference peak is defined as $M$.  (b) Model fitting (blue curve) of experimental profile (dark)  taken near the natural defect, with fitted damping ratio of 0.3. (c) Dependences of $M$ value and plasmon damping ratio $\gamma$ on ion dwell time. The red dashed line shows the trend of damping ratios for larger dwell times.}
\label{fig3}
\end{figure}

The discrepant near-field images for two  different ion doses in Fig. \ref{fig1} indicate that ion dose is a key factor that determines properties of plasmon reflection. Fig. \ref{fig3}(a) compares plasmon profiles of varied dwell times, where all near-field signals are normalized to that of graphene far from the reflectors. As shown, the appearances of plasmon interference profiles among dwell times of 100, 5 and 1 $\mu$s are different. Near-field signals of the first peak change with varied doses. The peak height of plasmon interference profiles, which is defined as visibility $M$, is an important parameter related to plasmon reflection from graphene boundary. For large $M$, plasmon reflection should be strong and additionally the damping of plasmon waves cannot be heavy. Moreover, varied damping tendencies imply different plasmon damping near the etched boundaries.

To extract plasmon damping information with different exposure dose, the near-field signals are fitted. The metallic tip launches plasmon waves propagating around with wave vector of $q=q_1+iq_2$, where the imaginary part $q_2$ stands for plasmon damping. The ratio between $q_2$ and $q_1$ defines plasmon damping ratio $\gamma=q_2/q_1$. Propagating plasmon waves reflect at ion etched boundaries then are scattered by the AFM tip and collected by detector. The field can be fitted as \cite{GBO2014,WLG2014}
\begin{align} \label{equ1}
{\Psi=\Psi_G+a\frac{\exp(iq*2x)}{\sqrt{x}}},
\end{align}
where $\Psi_G$ is the near-field contribution from graphene itself, $a$ is a complex parameter and $x$ is the distance from the detection position to the reflection boundary. In order to account for the finite size of the tip which generates a spatial averaged near-field response, the weight function $\Theta$ convolved with the spatially varying response $\Psi$ yields s-SNOM near-field signal \cite{GBO2014} $s(x)=(\Psi \ast\Theta)(x)$. $\Theta$ is adopted as Gaussian function peak at $x$ with a width of 10~nm.

Affected by inhomogeneous dopings, near-field signals near the first plasmon peaks vary unpredicted by Eq. \ref{equ1}. Thus signals away from the first peak were fitted. The fitted result for plasmon reflected from a natural boundary is shown in  Fig. \ref{fig3}(b), with the damping ratio valued $\gamma=0.28$ when the fitted curve and experimental profile show identical oscillation and damping tendency. Similar fittings are processed for ion dwell times not larger than 20 $\mu$s where the second plasmon peaks can be distinguished. The acquired damping ratios for different ion doses are plotted in  Fig. \ref{fig3}(c), together with the $M$ values. Moreover, the  damping ratios for larger dwell times can be deduced from the trend indicated by the red line. With decreased dwell times, as shown, plasmon waves decay weaker near etched boundaries, to a damping ratio of $\gamma=0.27$ when dwell time is 1 $\mu$s. This is nearly identical with that from natural graphene boundaries, implying little graphene degeneration occurs near the etched boundaries. The relation of decreased damping ratio with less ion dose is apparent, as fewer defects are introduced to graphene near etched boundaries at decreased ion dose.

Another parameter in Fig. \ref{fig3}(c) is the fringe visibility $M$, which experiences a rise than drop with more ions irradiating. The rise is easy to understand, as more ions strengthen the reflection boundaries with introducing more reflection components.  When the ion dose is large enough that the graphene layer is cutted into two pieces, the reflection ratio of plasmon waves should tend to saturate. Increasing ion dose can barely contribute to the reflection of plasmon waves, but introduces more defects near reflection boundary. This results in increased damping to plasmon waves propagating towards and away from the boundary, thus the drop of plasmon wave amplitude at interference peak.

The fixed damping ratios in fitting of  Fig. \ref{fig3} reveal averaged graphene qualities over the distance of one plasmon wavelength. The influence of ion etching on graphene further away is unrevealed. Former research on the spatial influence of ion beam on etched graphene relies on Raman spectrum characterization \cite{ABN2012,LZG2015,TVM2015}, which, restricted to diffraction limit, cannot reach the spatial resolution of smaller than 400 nm.  Here, the influence of ion etching can be  identified through near-field plasmon imaging.
\begin{figure}[htb]
\centerline{\includegraphics[width=8.5cm]{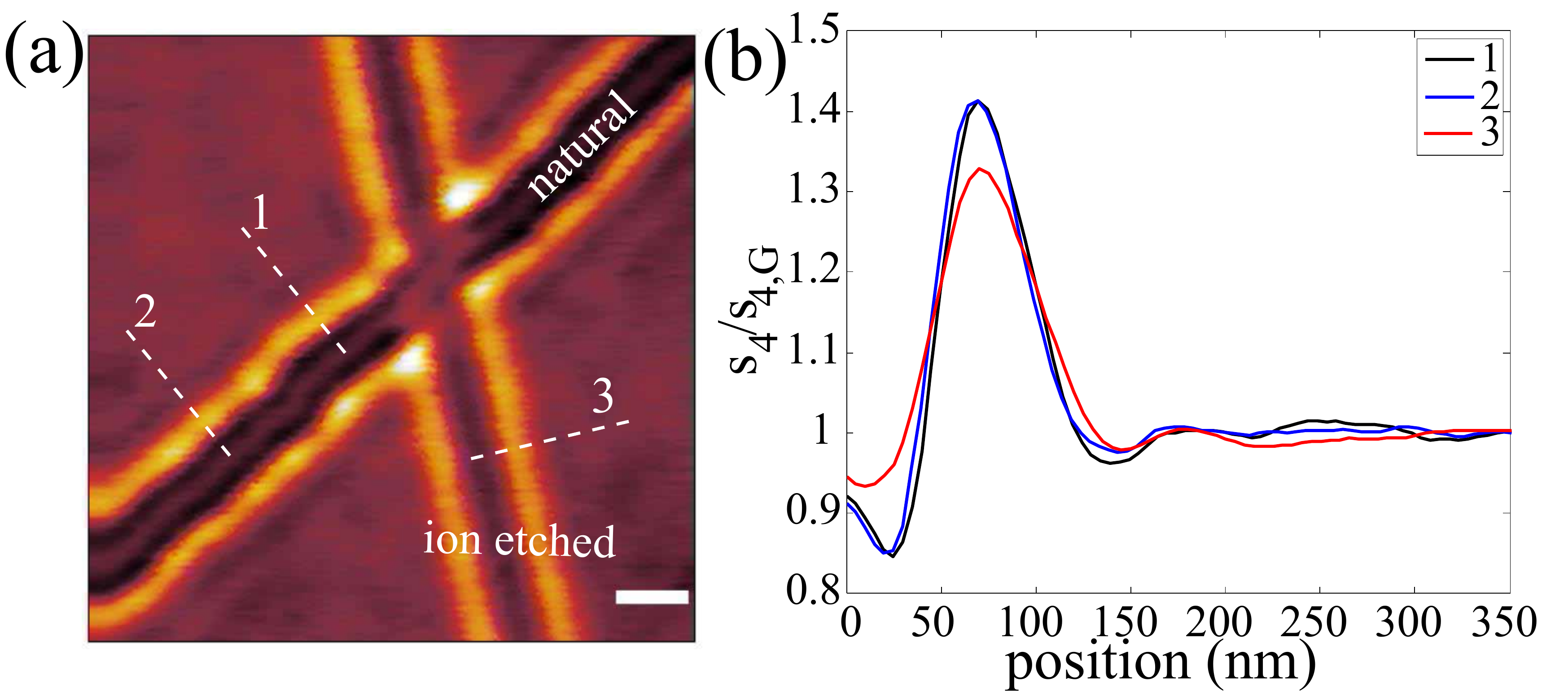}}
\caption{Influence of FIB etching on nearby graphene plasmon propagation. (a) Near-filed imaging of graphene with crossed FIB etched defect and natural defect. Incident wavelength is 10.616 $\mu$m.  Scale bar, 100 nm. (b) Line profiles taken along the dashed lines in (a). }
\label{fig4}
\end{figure}

Fig. \ref{fig4} shows the near-field image at the cross connection zone between a natural line defect and  the FIB induced defect at dwell time of 5 $\mu$s. Line profiles near the natural defect along lines of different distance to the ion etched area are plotted in Fig. \ref{fig4}(b). Line 2 is 420 nm away to the irradiated area and line 1 is about 170 nm away to that. As can be seen, the near-field signals exhibit almost coincident distributions, revealing that no depravation of graphene plasmon properties occur for 170 nm away from ion irradiated area. Therefore negligible change to graphene optical properties occur for separation larger than 170 nm far from ion etched zone,  guaranteeing
the availability of graphene away from the etched plasmon reflector in integrated applications. Moreover, profile taken along the ion etched boundary reveals a slightly larger damping compared to those from the natural one, in coincidence with the damping ratios shown in Fig. \ref{fig3}.

\begin{figure}[htb]
\centerline{\includegraphics[width=9.5cm]{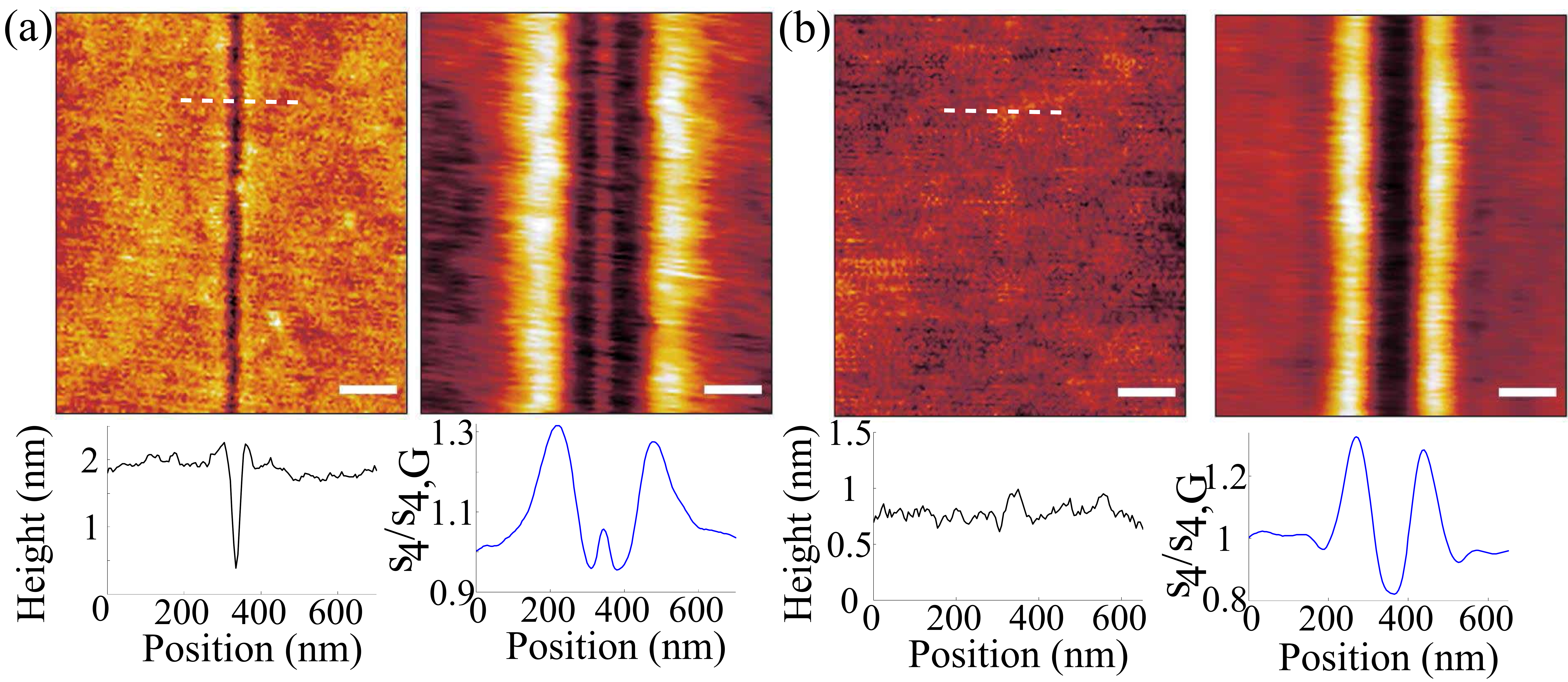}}
\caption{Nano-imaging of graphene with line defects fabricated by tilted ion beams. The ion beam is tilted 30 degree with respect to substrate's normal direction. (a), (b) AFM (left) and near-filed imaging (right) near FIB etched graphene. Incident wavelength is 10.653~$\mu$m. Dwell time is 100~$\mu$s (a) and 10 $\mu$s (b). Scale bars, 100~nm.}
\label{fig5}
\end{figure}

Our works above employ the configuration of perpendicular irradiation. Ion beam direction was proposed as an important factor for improving graphene modification quality \cite{LKK2011}. Here, ion beam is tilted 30$^\circ$, and corresponding near-field images are shown in Fig. \ref{fig5}. Graphene cutted zone of about 30 nm wide and 1.5 nm deep is also processed, while hillocks near etched zone are almost absent, suggesting that much fewer adsorbates exist.  Similar plasmon fringes are observed, as for large ion dose, there exists only the first plasmon peak, while for fewer dose, second interference peak emerges. Plasmon profile in Fig. \ref{fig5}(b) is fitted with Eq. \ref{equ1} and damping ratio is $\gamma$=0.38, smaller than 0.45 of the same dose while ion beam incidents perpendicularly, indicating less graphene defect exists in proximity to the reflection boundary.

\section{\label{sec:level6}CONCLUSION}
In conclusion, we demonstrate controllable plasmon reflection at graphene defect boundaries introduced by ion beams. Through near-field plasmon imaging, we observed plasmon interference fringes near ion etched boundaries. The acquired plasmon dispersion at varied frequencies verifies efficient plasmon reflection near ion etched graphene zone. For varied ion doses, different plasmon reflection properties are obtained. Large dose results in increased plasmon damping while decreasing ion dose increases the plasmon propagating distance,  and nature like graphene defect can be achieved. Moreover, the reflection ratio of plasmon waves can also be controlled by changing the ion dose. This ion dose related reflection can be applied in fabrication of plasmon elements with varied characteristics. A simple graphene wedge-shape structure was fabricated and exhibits excellent plasmon properties. The influence area of ion beam are also measured, revealing that for distances larger than 170 nm, no degeneration of graphene plasmon properties can be observed, which guarantees the applications in integrated devices. To obtain less plasmon damping, ion etching at tilted angle is demonstrated as an efficient method. Improved plasmon properties should be achieved with He$^+$ which causes reduced proximity effect \cite{BLS2009,LBW2009}. Our study shows that FIB is an simple and efficient tool for engineering graphene functional plasmon elements of varied properties, which is promising for applications in integrated photonic devices.

\begin{acknowledgments}
This work was financially supported by the National Basic Research Program of China (2013CB328702), Program for Changjiang Scholars and Innovative Research Team in University (IRT0149), the National Natural Science Foundation of China (11374006) and the 111 Project (B07013). We thank the Nanofabrication Platform of Nankai University for fabricating samples, and support from Synergetic Innovation Center of Chemical Science and Engineering (Tianjin) and Collaborative Innovation Center of Extreme Optics (Shanxi).
\end{acknowledgments}

%
%

\end{document}